\documentclass[12pt,a4paper]{article}
\usepackage{amsmath,amsfonts,amsthm,graphicx,lscape}
\usepackage[dvips]{psfrag}
\usepackage{cite}

\usepackage[normalem]{ulem}

\usepackage{textcomp}
\usepackage{amsmath,amsthm,amssymb,mathrsfs,cases}
\usepackage{amsfonts,graphicx,lscape}

\numberwithin{equation}{section}

\def\beqa{\begin{eqnarray}}
\def\enqa{\end{eqnarray}}
\def\beq{\begin{equation}}
\def\enq{\end{equation}}

\begin{document}
\title{
\vspace{-9mm}
Comment on 
``A counterpart of the WKI soliton hierarchy associated with 
\mbox{$\mathrm{so}(3,\mathbb{R})$}"
}
\author{Takayuki \textsc{Tsuchida}
}
\maketitle

In the recent paper 
(Wen-Xiu Ma, Solomon Manukure and Hong-Chan Zheng, 
arXiv:1405.1089), 
the authors 
proposed an integrable hierarchy 
different from 
the well-known Wadati--Konno--Ichikawa 
(WKI)
hierarchy. However, 
%
using a simple linear change of dependent variables such as 
\[
p + \mathrm{i} q  =: u , \hspace{5mm} p - \mathrm{i} q=:v, 
\]
one can check that 
their hierarchy is 
equivalent 
to the WKI 
hierarchy. 
Note, for example, that 
the second nontrivial flow in their hierarchy 
((3.18) in arXiv:1405.1089), 
\begin{align}
\left[
\begin{array}{c}
 p
\\
 q
\\
\end{array}
\right]_{t_1} = - 
\left[
\begin{array}{c}
\left( \frac{p_x}{(p^2 + q^2 +1)^{\frac{3}{2}}} \right)_{xx}
\\[4mm]
 \left( \frac{q_x}{(p^2 + q^2 +1)^{\frac{3}{2}}} \right)_{xx}
\\
\end{array}
\right], 
\nonumber 
\end{align}
is 
transformed to 
the corresponding flow in the WKI 
hierarchy: 
\begin{align}
\left[
\begin{array}{c}
 u
\\
 v
\\
\end{array}
\right]_{t_1} = - 
\left[
\begin{array}{c}
\left( \frac{u_x}{(1+uv)^{\frac{3}{2}}} \right)_{xx}
\\[4mm]
 \left( \frac{v_x}{(1+uv)^{\frac{3}{2}}} \right)_{xx}
\\
\end{array}
\right]. 
\nonumber
\end{align}

For the same reason, 
some 
``new integrable" hierarchies  
proposed by Wen-Xiu Ma and coworkers in recent e-prints 
are equivalent to the already 
known ones. 

\end{document}